\documentclass[twocolumn,prl,aps,amsmath,amssymb,color,superscriptaddress]{revtex4}
\usepackage{stmaryrd}
\usepackage{amsmath}
\usepackage{amssymb}
\usepackage{graphicx}
\usepackage{dcolumn}
\usepackage{bm}
\usepackage{tabularx}
\usepackage{color}
\begin{document}	
	\begin{titlepage}
		\title{A Universal Description of Workfunction}
		\author{Zeyu Jiang}
		\author{Damien West}
		\author{Shengbai Zhang}
		\email{zhangs9@rpi.edu}
		\affiliation{Department of Physics, Applied Physics and Astronomy, Rensselaer Polytechnic Institute, Troy, NY, 12180, USA}
		\date{\today}
		\begin{abstract}
		At the surfaces of materials, the bulk symmetry of the charge density is broken and electron spill-out into the vacuum region creates a surface dipole. Such spill-out has been historically calculated by Lang and Kohn [Phys. Rev. B \textbf{3}, 1215 (1971)] using average electron density to sucessfully explain the workfunction in metals. However, despite its initial success, in the fifty years since it has not been extended beyond simple metals. Here we show that the degree of charge spill-out is largely controlled by the innate bulk workfunction $\phi_I$, which is the Fermi level position of $\it bulk$ relative to the ideal vacuum. By incorporating the contribution of $\phi_I$ to the surface dipole we show that Lang-Kohn's $\it jellium$ based approach can be broadly expanded to understand the workfunction over a wide range of metals, semiconductors, and insulators. 
				
		
	      

		\end{abstract}
		\maketitle
		\draft
		\vspace{2mm}
	\end{titlepage}
	
	Workfunction is a key part of the original theory of Einstein on photoelectric effect\cite{Einstein1905}. Its importance to science spans a wide range of topics from thermionic emission and energy conversion\cite{Huffman2003}, chemisorption\cite{Chan1996}, surface reconstruction\cite{Chan1996}, surface chemistry\cite{Bare2003} and chemical sensors\cite{Bergveld1998}, material’s fracture toughness\cite{Hua2016} and mechanical strength\cite{Lu2016}, to free-electron lasers\cite{Jensen2003}, to name a few. In the context of the so-called Schottky-Mott limit\cite{Schottky1939,Mott1939} and Anderson rule\cite{Anderson1960}, workfunction also serves as a rough guide to band alignment for electronic device design\cite{Kroemer2000,Kahn2016} and interfacial diagnosis\cite{Li2017}, and to match redox potentials for photoelectrochemical reaction and energy conversion\cite{Sivula2016}. On the theoretical side, density functional theory (DFT) based methods provide reliable access to accurate results of workfunctions\cite{Waele2016}, using either a periodic supercell with ``slab plus vacuum" geometry or an open-boundary real-space approach. However, the physics and factors that control the workfunction remain poorly understood, despite a century of intense efforts. 
	
	This happens in part because most of our understanding on solids are based on the construct of infinitely large bulk, thanks to the Bloch theorem and to the wide adoption of Fast Fourier transform (FFT) algorithms. In this regard, the surface of solid is viewed as the interface between an infinitely large bulk and an infinitely large ``empty" space as the vacuum, and the workfunction corresponds to the relaxation of this interface. Yet, one always needs a common reference energy to quantify interactions and align electronic levels between subsystems when placed together. For example, in quantum chemistry, the reference energy of a molecule is chosen as the potential energy at position infinitely far away from the molecule. An obvious drawback of the infinitely large solid model is that the reference energy gets ``lost", as there is no ``infinitely far" position for an infinitely large bulk. Without identifying the reference energy for periodic bulk, we don't know how to place an infinitely large solid to the vacuum, and a clear-cut description of workfunction would be completely impossible.
	
	Recently, we identified the reference energy of infinitely large solid as the \emph{ideal} vacuum level\cite{Choe2018}. It is important to note that, although the macroscopic monopole and dipole vanish in a periodic bulk, due to charge neutrality and translational symmetry, the inhomogeneous distribution of bulk charge still provides a finite electric quadrupole tensor, $\textbf{Q}$. Furthermore, across the surface of an ideal (truncated) bulk crystal with a given orientation $\hat{\textbf{n}}$, $\textbf{Q}$ is related to the offset of the bulk average electrostatic potential $\overline{V}$ from the ideal vacuum level $V_{ideal}^{\hat{\textbf{n}}}$ by, \cite{Choe2021}
	\begin{equation}
		V_{\textbf{Q}}^{\hat{\textbf{n}}} = V_{ideal}^{\hat{\textbf{n}}} - \overline{V} = \frac{4\pi}{\Omega} \hat{\textbf{n}}^{T} \overleftrightarrow{\textbf{Q}} \hat{\textbf{n}},
	\end{equation}
	where $\Omega$ is the unit cell volume. In standard bulk calculations, $\overline{V}$ is often used as the reference energy, i.e., $\overline{V}=0$. As such, all bulk energy levels are orientation independent. In contrast, now the vacuum level becomes an orientation dependent quantity.
	
	In this work, we formulate a theory of workfunction and show that the bulk quadrupole in Eq. (1) plays a critical role in determining the overall workfunction. As a real system will always involve surface relaxation, in particular, the electronic relaxation in our theory, the actual workfunction must contain two parts: a pure bulk term $\phi_{I}$ and a surface-relaxation dipole term $V_{D_R}$. By a real-space charge and potential analysis, we show that to a first-order approximation, $V_{D_R}$ can be evaluated using a uniform background charge for ions in the solid region, provided that the effect of $\phi_{I}$ is properly incorporated. The results are compared with first-principles slab calculations for 21 real solids. Good agreement is found across board from metals, semiconductors, to insulators, which establishes the physical origin of workfunction.
	
	
	In general, the workfunction $\phi^{\hat{\textbf{n}}}$ of a solid is a direction dependent (${\hat{\textbf{n}}}$) quantity defined by the difference between the vacuum potential $V_{0}^{\hat{\textbf{n}}}$ in the near-surface region and bulk Fermi level $E_F$. Here, we decompose it into a pure surface term plus a pure bulk term as follows
	\begin{equation}
		\begin{aligned}
			\phi^{\hat{\textbf{n}}} &= V_{0}^{\hat{\textbf{n}}} - E_{F} \\ 
			& = \left( V_{0}^{\hat{\textbf{n}}} - V_{ideal}^{\hat{\textbf{n}}} \right)  + \left(V_{ideal}^{\hat{\textbf{n}}} - E_{F}\right) \\
			& = V_{D_R}^{\hat{\textbf{n}}} + \phi_{I}^{\hat{\textbf{n}}},
		\end{aligned}
	\end{equation}
	where $V_{D_R}^{\hat{\textbf{n}}}$ is the surface relaxation dipole potential and $\phi_{I}^{\hat{\textbf{n}}}$ is the \emph{innate} workfunction, as a pure bulk contribution to $\phi^{\hat{\textbf{n}}}$. Using Eq. (1), we can further write $\phi_{I}^{\hat{\textbf{n}}}$ in Eq. (2) as
	\begin{equation}
		\phi_{I}^{\hat{\textbf{n}}} = V_{ideal}^{\hat{\textbf{n}}} - E_{F} = V_{\textbf{Q}}^{\hat{\textbf{n}}} - \left(E_{F}-\overline{V}\right).
	\end{equation}
	Equation (2) may be contrasted to first-principles calculations which, on one hand, yield accurate $\phi^{\hat{\textbf{n}}}$ but do not delineate bulk and surface contributions, and model studies\cite{Halas1998,Wong2003,Gutierrez2007,Brodie2014,Tran2019} which usually rely on {\it ad hoc} assumptions and numerical fitting. 
	
    \begin{figure}[tbp]
    \includegraphics[width=\columnwidth]{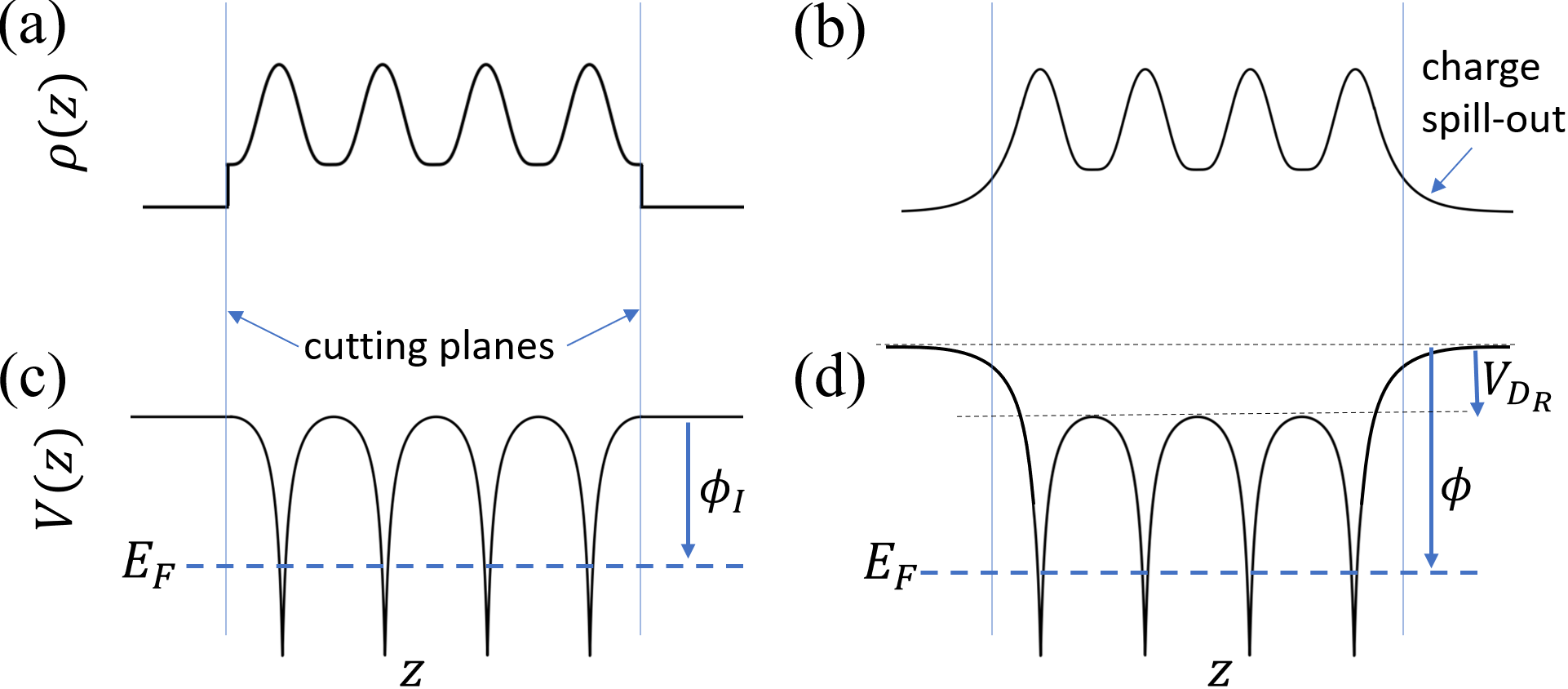}
    \caption{\label{fig:fig1} (Color online) Schematic of the charge density of a truncated slab, with the cutting planes depicted as verticle lines, before (a) and after (b) electronic relaxation. The electrostatic potential corresponding to the charge densities in (a) and (b) are shown in (c) and (d), respectively.}
    \end{figure}
	
	As has been shown in Ref. [18],  $\phi_I^{\hat{\textbf{n}}}$  is related to the bulk quadrupole, through $V_{ideal}^{\hat{\textbf{n}}}$, and can be straightforwardly calculated via a unit cell in DFT. $D_R^{\hat{\textbf{n}}}$, however, is physically determined by the extent to which the self-consistent charge spills out into the vacuum region at a particular surface. As illustrated in Fig. 1, the truncated wavefunction associated with an electron at the Fermi level would exponentially decay into the vacuum region, with a characteristic length which would depend on the difference in energies between the electronic state in the material and in the vacuum region. Before relaxation, this energy difference is simply $\phi_I^{\hat{\textbf{n}}}$. However, as charge spills out, the resulting surface dipole alters the alignment of the aforementioned states, yielding $\phi^{\hat{\textbf{n}}}$ , which needs to be determined self-consistently. 
	
	While such self-consistent calculations can be carried out for any specific material with slab supercell, yielding $\phi^{\hat{\textbf{n}}}$, the knowledge is not transferrable and provides little insight into fundamental questions about the nature of the spill-out or the workfunction. Within the context of DFT calculation, the non-interacting electrons experience the same effective potential, 
	\begin{equation}
	V_T(\textbf{r})=V_{ion}(\textbf{r})+V_{H}(\textbf{r})+v_{xc}(\textbf{r}),
	\end{equation}
	where the individual terms to the right correspond to the ionic electrostatic potential, the Hartree potential of electrons, and the exchange-correlation potential, respectively. In 1971, substantial progress was made in a more general understanding of the workfunction, where Lang and Kohn\cite{Lang1971} (LK) uncovered that details of the ionic positions could be abstracted away and that the workfunction of simple metals are well described through a $\it jellium$ approximation, in which the ionic charge is assumed to be uniformly distributed. While this approach has proved unsucessful for semiconductors, it provides a picture of metals wherein the workfunction can be simply understood by the average electron density of the material.
	
    \begin{figure}[tbp]
    \includegraphics[width=\columnwidth]{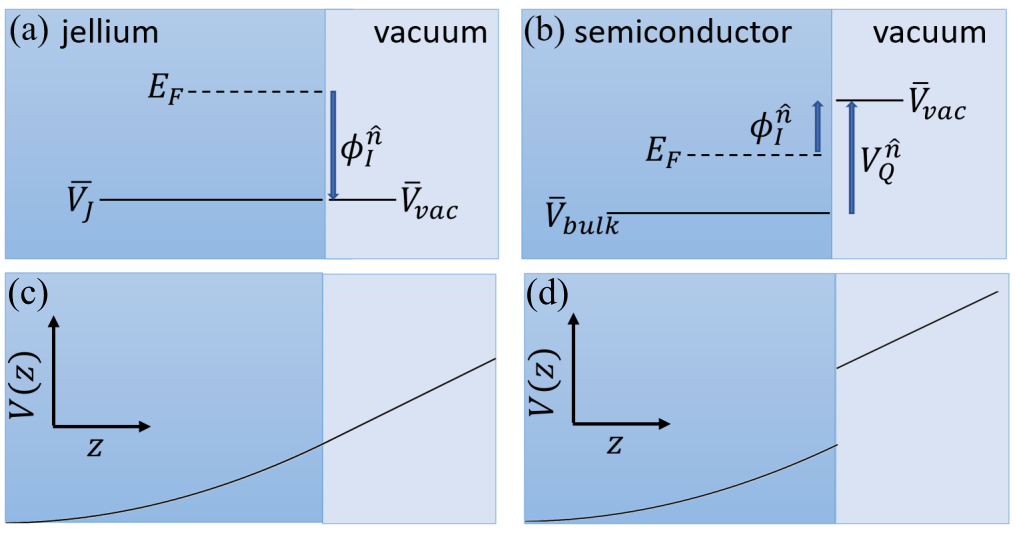}
    \caption{\label{fig:fig2} (Color online) Schematic of the electrostatics of a truncated bulk. The alignment of averaged total electrostatic potential between bulk and vacuum region is shown for (a) jellium and (b) semiconductor. (c) Ionic potential of jellium. (d) The first-order approximated ionic potential of semiconductor, where the ionic charge is replaced by jellium but the bulk electric quadrupole induced offset is preserved.}
    \end{figure}	

	Fig. 2(a) summarizes the electrostatics of LK's approach. The uniform charge distribution leads to a continuous translational symmetry breaking only in the direction perpendicular to surface, simplifying the system into a one-dimensional model. The potential due to the ionic $\it jellium$ is shown in Fig. 2(c), where the constant charge density of the bulk region leads to a quadratically increasing potential which becomes linear in the vacuum region. For such a truncated $\it jellium$, the total electrostatic potential in the bulk region is the same as the vacuum and $E_F$ is typically higher than the vacuum level, yielding a negative $\phi_I^{LK}$. We note here that $\phi_I$ is the driving force for the charge spill-out, and a very negative $\phi_I$ would have significant charge spill-out while a very positive $\phi_I$ would have minimal charge spill-out, yielding $\phi$ which is very different or similar to $\phi_I$, respectively. Hence, whatever simplification to the problem used, accurate representation of $\phi_I$ is necessary to obtain the approriate physics. 
	
	In this context, LK's wide sucess for metals and failures for semiconductors/insulators can be understood. In particular, as systems become more metallic the details of the ionic coordinates are screened away and the electric quadrupole of the bulk system approaches zero. Hence, $\phi_I$ is well represented in a $\it jellium$ approximation. For semiconductors on the other hand, as depicted in Fig. 2(b), the incomplete screening leads to significant charge inhomogenity resulting in an electric quadrupole which yields an offset $V_Q^{\hat{\textbf{n}}}$ in the average potential of bulk relative to vacuum. 
	
	As the charge spill-out is in the $z$-direction, it is expected to be quite insensitive to inhomengeties in the $xy$-plane and hence well described within the $\it jellium$ appoximation. However, it will still be sensitive to inhomogenity of charge in the $z$-direction. In order to capture this essential physics of the workfunction, we incorporate the discontinuity in the potential associated with the electric quadrupole as a step function in the $z$-direction as shown in Fig. 2(d). In addition, the difference in the exchange-correlation potential between $\it jellium$ and the real solid is reflected by the difference in $E_F$. By adding a second step function to the interface, the $E_F$ of $\it jellium$ relative to $\bar V_{vac}$ (i.e., $\phi_I$) can be well represented. In this model, charge spill-out is still calculated within the $\it jellium$ method but the potential at the interface is modified such that $\phi_I$ of $\it jellium$ is corrected, hence we refer to it as the $\phi_I$ corrected $\it jellium$ ($\phi_IJ$) model.

	To determine how well the relevant physics of the surface relaxation, $D_R$, can be captured with such a simple $\it jellium$ approximation in the effective potential outlined above, we perform calculations on a series of 21 materials including face-centered-cubic (fcc) structure elemental metals (Li, Na, K, Be, Mg, Ca, Al, Ga), diamond/zinc-blende structure semiconductors (Si, BN, BeO, AlP, GaAs, AlSb) and rock-salt structure insulators (LiF, LiBr, NaF, NaCl, KCl, MgO, CaO). Our real-crystal calculations were performed using density functional theory (DFT) by VASP package\cite{Kresse1996} and exchange-correlation potential of local density approximation (LDA)\cite{Ceperley1980}, with interactions bewteen ion cores and valence electrons described by the projector augmented-wave (PAW) method\cite{PAW}. Our $\it jellium$ calculations were carried out by GPAW package\cite{Enkovaara2010}. We used the optimized lattice structures from the Material Project Database\cite{Jain2013} and fixed atomic positions for all the calculations. To simplify the discussion, we consider in this work only the non-polar surfaces for binary materials.
	

    \begin{figure}[tbp]
    \includegraphics[width=1.0\columnwidth]{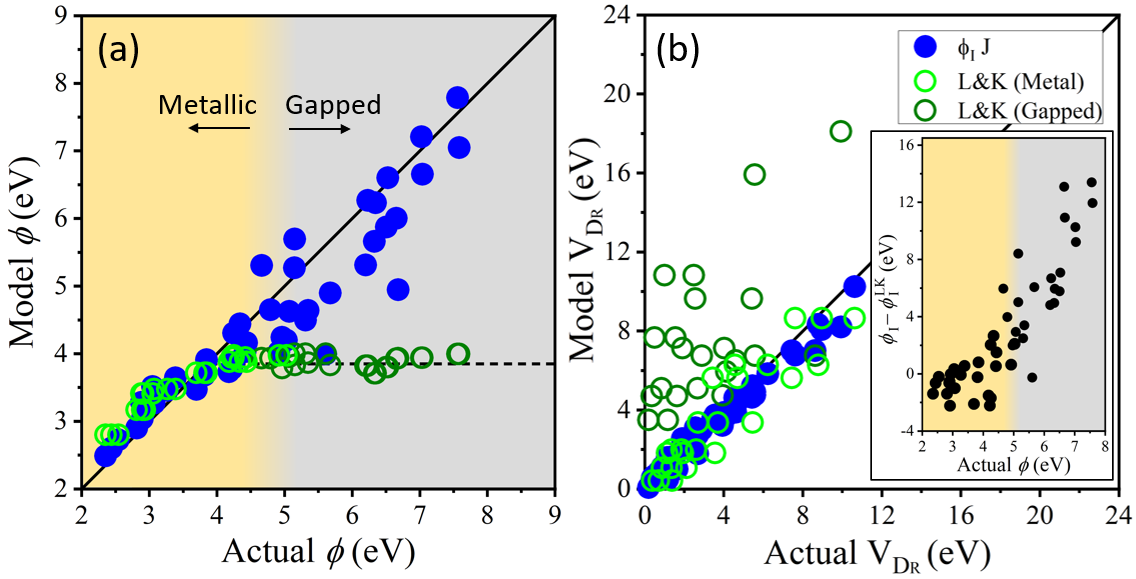}
    \caption{\label{fig_res} (Color online) Calculated $\phi$ (a), $V_{D_R}$ (b), and $\phi_I - \phi_I^{LK}$ (inset of (b)) for 21 metallic or gapped systems. The horizontal axis (Actual) indicates the values obtained from fully atomistic DFT calculations. The vertical axis shows results of our $\phi_IJ$ model (blue shaded).The results from Lang and Kohn's approach for metals (light-green open) and non-metals (dark-green open) are included as reference. Most materials are considered with more than one surface orientations, thus the data points are more than 21.}
    \end{figure}


    The results are summarized in Fig. \ref{fig_res} with comparison to the real-crystal calculations. Here the horizontal axis corresponds to fully atomistic DFT calculation of the crystal (Actual), while the vertical axis shows the values calculated from our $\phi_IJ$ model as well as LK's approach (Model). In Fig. \ref{fig_res} (a) it can be seen that the $\phi_IJ$ model largely follows the diagonal line (corresponding to model $\phi$ = actual $\phi$), with some spread. The largest deviations are found for semiconductors, in which the deviation can exceed 1 eV in the case of the (100) and (110) surfaces of Si and BN, respectively. For reference, the results from LK's approach are shown by the open symbols. While both methods work equally well for metallic systems, as $\phi$ becomes larger, LK's approach predicts instead an almost constant $\phi$ of nearly 4 eV. Note that larger $\phi$ is also associated with the system developing an increasing bandgap. As the gap becomes very large, $\phi_{I}$ becomes critically important and the contribution of $D_{R}$ can be understood as secondary response to $\phi_{I}$. 
  
 In Fig. \ref{fig_res} (b) we directly compare this electron spill-out within the $\phi_IJ$ model and LK's approach. 
While both using $\it jellium$ to calculate the electron spill-out and the corresponding surface dipole which develops, the $\phi_IJ$ model corrects the value of $\phi_I$ within the $\it jellium$ calculation so that the Fermi level position relative to vacuum is well represented before considering the charge spill-out. Here it can be seen that while the $\phi_IJ$ model largely reproduces the actual $V_{D_R}$ across the range of metals, semiconductors, and insulators, LK's approach for nonmetals (dark green open circles) generally leads to substantial overestimation.

 In order to understand the general agreement of the two approaches for the case of $\it metals$, we examine the difference of $\phi_{I}$ in the two approaches, $\Delta\phi_I=\phi_{I}-\phi_{I}^{LK}$, shown as an inset in \ref{fig_res} (b) (note here that $\phi_I^{LK}$ is simply $E_F$ relative to the average potential of $\it jellium$, as depicted in Fig. 2 (a)).
From the inset, $\Delta\phi_I$ becomes large in gapped systems confirming the significant underestimation of $\phi_{I}$ in LK's approach, which explains why it fails for semiconductors and insulators. In contrast, $\Delta\phi_I$ is much smaller for metallic systems, thus both approaches can work equally-well in this range. The influence of ignoring $\phi_I$ is much weaker in metallic systems so the relaxation dipole $D_{R}$ dominates, however, in gapped systems both $\phi_{I}$ and $D_{R}$ are non-negligible.

 These results highlight the essential role of the electric quadrupole in the understanding of workfunction. Electric qaudrupole is the direct reflection of charge inhomogeneity under broken continuous translational symmetry, which manifests more intensely in semiconducting and insulating materials with highly localized bonds. Although all the direct structural information is removed in our $\it jellium$ calculation, this effect of inhomegenity is preserved by adequately accounting for $\phi_I$. Hence, the workfunction of both metallic and gapped systems can be understood from jellium method on the same footing.

    \begin{figure}[tbp]
    \includegraphics[width=1.0\columnwidth]{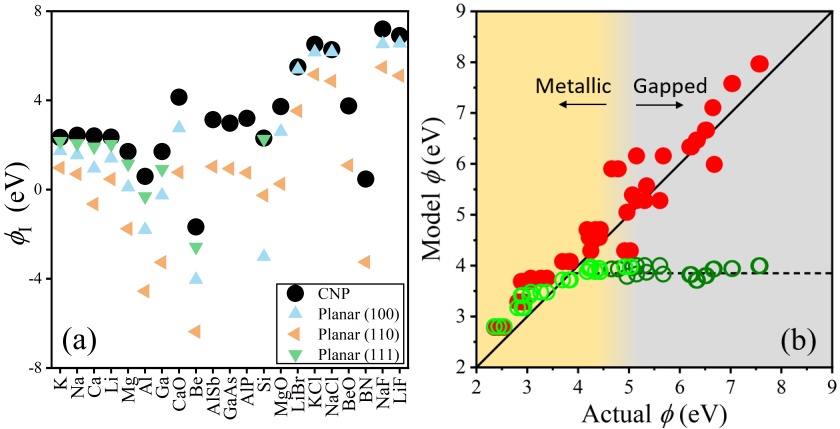}
    \caption{\label{fig_phicomp} (Color online)  (a) Comparison of $\phi_I$ calculated using CNP cut (circles) and planar cut in different directions (triangles). The materials are ordered by ascending actual $\phi$.(b) Same as Fig. 3(a), but the results of $\phi_IJ$ model (red shaded) are calculated using $\phi_I$ obtained from CNP cut. The result of Lang and Kohn's approach (light/dark-green open) is same as that in Fig. 3(a).}
    \end{figure}

	Despite the success of the above description, we stress that $\phi_I$ and the associated $D_R$ are highly dependent on the orientation of the terminating surface, with $\phi_I$ varying over 5 eV. While this direction dependence largely vanishes for the workfunction, $\phi$, within $\phi_IJ$ model, consistent with the experimental observation (which typically exhibits direction dependence of less than 1 eV \cite{Tran2019}), it is nevertheless disconcerting given that it seems rather unphysical. The origin of this large deviation in $\phi_I$ can be traced back to our choice of a simple planar truncation of the surface. Depending on the orientation of the surface, such a cut can easily cut very close to the core of the atoms. Physically, we know that the charges should maintain both the translational symmetry (which has been preserved in our procedure) and rotational symmetry (or point group) of the crystal (which is violated by a planar cut). As such, a large part of the calculated surface relaxation is to restore the approximate rotational symmetry of bulk charge rather than relaxing with respect to the vacuum.
	

	We note, however, a planar cut is not a fundamental limitation of our theory. A 3D cut to create the surface would serve the same purpose, as long as the monopole and dipole terms for the bulk unit cell are kept zero. In such a configuration, the periodic cell representing the bulk charge density has a shape distinct from the parallelepiped associated with the Bravis vectors. One such construction is the charge-neutral polyhedron (CNP)\cite{Tung2016}, which preserves both the translational and rotational symmetry of the bulk crystal. In particular, a CNP is defined by partitioning the charge density into atom-centered charge-neutral polyhedra in a way similar to the determination of the Wigner-Seitz cell. One cuts real space with planes perpendicular to bonds between the target atom and all its neighbors with the distance to the plane determined such that the total charge enclosed in the smallest polyhedron vanishes for each atomic specie. For commonly-studied elemental or binary materials, the CNP cell can be uniquely defined. 
	
	An interesting quality of the CNP partitioning for the materials studied here is that the innate workfunction, $\phi_{I}^{CNP}$, is completely $\it independent$ of surface orientation and hence truly only a property of bulk. Comparison of $\phi_I$ using the planar and CNP cut are shown in Fig. 4(a). Furthermore, as $\phi_I^{CNP}$ is consistently larger than $\phi_I^{planar}$, it is associated with a smaller $D_R$ as it avoids cutting the charge density near the core region. Note that among the planar-cut directions, the (111) direction, which bisects nearest neighbor bonds, yield results most similar to the CNP method. Despite the substantial differences in the cutting methods, they yield quite similar results for the workfunction when used in the $\phi_IJ$ model, as shown in Fig. 4(b). While there appears to be a mild systematic over estimation of $\phi$ using the CNP cut, the results of semiconductors have been greatly improved. As $\phi_I^{CNP}$ is orientation independent, this cut naturally yields an orientation independent $\phi$, which, while only approximately true in experiment, is foundational in the chemical understanding of redox reactions and hints that the construction may have a deeper physical significance. 
	
	
	In summary, we have developed a unified understanding of the workfunction of solids, in which the bulk and surface contributions can be conceptually separated and studied independently. Similar to the pioneering work of Lang and Kohn on metals, we find that charge spill-out at the surface of materials can be well described by a simple $\it jellium$ approximation. By correcting the Fermi level position of $\it jellium$ relative to vaccum to match that of the real crystal, $\phi_I$, (determined from the bulk electric quadrupole) we significantly extend the approach and show that it can yield accurate workfunctions over the entire range of metals, semiconductors, and insulators. As the bulk-vacuum interface is an extreme example of a material interface, it would be natural to expect the dipoles at heterostructures to be considerably smaller than those found here. This may explain the sucess of simply using $\phi_I$ to explain the valence band offsets of semiconductors \cite{Choe2021}. 


	\begin{acknowledgments}
		S. Z. thanks Duk-Hyun Choe for fruitful discussions. This work was supported by the U.S. DOE Grant No. DE-SC0002623. The supercomputer time sponsored by National Energy Research Scientific Center (NERSC) under DOE Contract No. DE-AC02-05CH11231 and the Center for Computational Innovations (CCI) at RPI are also acknowledged.  
	\end{acknowledgments}

\end{document}